\documentclass[%
 reprint,
 amsmath,amssymb,
 aps,
floatfix,
]{revtex4-2}

\usepackage{tabularx}
\newcolumntype{Y}{>{\centering\arraybackslash}X}

\usepackage{makecell}
\usepackage{graphicx}
\usepackage{dcolumn}
\usepackage{bm}

\begin{document}

\preprint{APS/123-QED}

\title{Renormalization Group Approach to Percolation in Hierarchical Lattices}

\author{Abe Levitan}
\email{alevitan@mit.edu}
\affiliation{Department of Physics, Massachusetts Institute of Technology, Cambridge MA 02139}

\date{\today}

\begin{abstract}
Percolation refers to an interesting class of problems related to the properties of disordered systems, usually formulated in terms of objects randomly placed on an underlying lattice or continuum. Despite the simplicity of the setup, most percolative systems undergo a phase transition from a disconnected state with many disjoint clusters to a state where a finite fraction of the lattice sites are connected to a single cluster. As in the case of thermodynamic phase transitions, power law dependencies generically near the critical percolation threshold. The origin of these dependencies can be understood through the lens of scaling and renormalization, and indeed many quantitative results can be acquired using these tools. In this paper we study the percolation problem on a hierarchical lattice, where exact results for the critical exponents can be obtained from a decimation procedure. We calculate analytic results for the full set of geometric critical exponents and confirm their consistency with simulation. Finally, we set up an interesting renormalization group for the conductivity of the system and use it to computationally extract the conductivity exponent $t$.
\end{abstract}

\maketitle

\section{Bond Percolation}

Bond percolation refers to the class of problems related to the placement of bonds on a discrete lattice. A specific bond percolation problem is defined by specifying a lattice of bonds connected by vertices, and a probability $p$ that each bond will exist. A variety of questions can then be asked about the system at a particular value of $p$, such as: Does the size of any cluster reach $\infty$? If such a cluster exists, what is its density? What is its conductivity?

All percolation problems have a critical percolation threshold $p_c \in [0,1]$ at which a spanning cluster first appears. Near that critical threshold, it is possible to get universal answers to the questions above in the form of scaling exponents, which are generally valid for classes of lattices grouped by dimensionality $d$ or some other distinguishing characteristic. Before exploring results on a particular lattice, we will define the various scaling quantities and exponents, using a convention loosely based on \cite{stauffer} and \cite{percflow}.

The first quantity to define is the order parameter $P(p)$, defined as the density of the spanning cluster. This goes to zero in the disordered phase when $p<p_c$, and scales like $(p-p_c)^\beta$ when $p>p_c$. Related to this quantity are the cluster statistics $n(s,p)$, defined as the number of clusters of size $s$ per bond in the underlying lattice. It is immediately apparent that $P(p) + \sum_s s n(s,p) = 1$.

At $p_c$, we generically expect the cluster statistics to fall off with a power law in $s$ (as will be confirmed for the specific case we study). As $p$ moves away from $p_c$, we expect a size scale (equivalently, a mass scale) to appear at which the power law behavior gives way to exponential decay. Therefore we define the exponents $\tau$ and $\sigma$ such that

\begin{equation}
n(s,p) \propto s^{-\tau} \exp\left(s (p-p_c)^\frac{1}{\sigma}\right)
\end{equation}

It is common to define the scaling of the mean number of clusters as $\sum_s n(s,p) \propto (p-p_c)^{2-\alpha}$ and the scaling of the mean cluster size as $\sum_s s^2 n(s,p) \propto (p-p_c)^{-\gamma}$, where the suggestive choice of exponent labels is due to connections with thermodynamic systems.\cite{percflow}

The final commonly discussed quantity is the correlation length $\xi(p)$, defined as the mean length of a cluster on the lattice. The divergence of this length scale at $p_c$ is characterized by the exponent $\nu$, $\xi(p) \propto (p-p_c)^{-\nu}$. In addition, in this work we specifically deal with the scaling of the conductivity $\sigma(p)$, traditionally taken to scale with a critical exponent $t$ such that $\sigma(p) \propto (p-p_c)^{-t}$.

\section{Hierarchical Lattices}

\begin{figure}
\includegraphics[width=\linewidth]{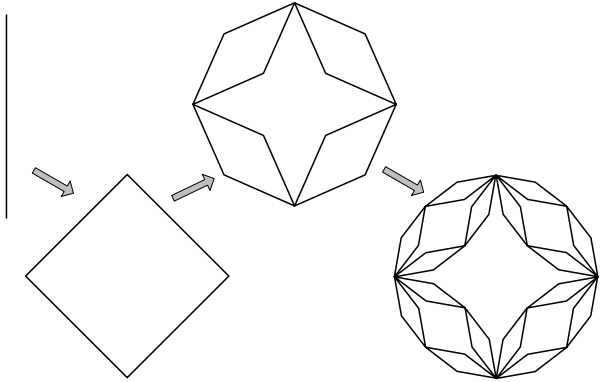}
\caption{The first 4 iterations in the generation of a hierarchical diamond lattice with effective dimension 2. Note that due to the high connectivity of certain points, it's reasonable to expect qualitatively different behavior in this lattice as compared to other 2D lattices.} \label{fig:berkerlattice}
\end{figure}

In this paper, we will study an example of a hierarchical lattice. These lattices can be generated by applying a replacement rule repeatedly, iteratively replacing a bond with a structural motif made of bonds. The lattice we use in this project is a hierarchical diamond lattice with $d_e = 2$ \cite{angulo93} \cite{roux91} and is shown in Figure \ref{fig:berkerlattice}. This system has been used in a variety of contexts as a crude approximation to the behavior expected in 2D lattices, although it is known that the critical exponents differ substantially in this system as compared to ``standard'' 2D lattices.

The advantage of working on a hierarchical lattice is that it will admit an exact renormalization group approach, and it is therefore possible to calculate the percolation threshold and critical exponents without approximations. While one can argue that such results are of limited utility in describing real systems, they are certainly worthwhile in a pedagogical sense because they lead to exploration of interesting variations on the renormalization-group approach, without introducing worries about the inexactness of the approximations that are required on more physical lattices.

\section{Geometric Exponents}

In this section we will calculate the full set of geometric exponents on the $d_e = 2$ hierarchical diamond lattice. The derivations of $p_c$ and $\nu$ loosely follow \cite{roux91} and \cite{young75}, while the calculation of the fractal dimension of the percolating cluster $d_f$ was independently developed (although it is possible that this method, or an equivalent one, has been used in the past in the literature).

It is apparent that we can proceed by decimation, replacing all the bonds single diamond motif with a coarse-grained bond. We note that the probability for a connected path to exist across a single diamond is $2p^2 - p^4$, which is the probability for a path to exist on either side minus the overcounting from the possibility of both paths existing at the same time. Under this coarse-graining step, therefore, we have:

\begin{equation}
p' = 2p^2 - p^4
\end{equation}

This mapping has 3 physical stationary points - stable points at $p=0$ and $p=1$, and an unstable critical point at

\begin{equation}
p_c = \frac{\sqrt{5} - 1}{2} \approx 0.618034
\end{equation}

Linearizing about the critical point using $p = \delta_p + p_c$, we find

\begin{equation}
\delta_p' = 4 \left(p_c - p_c^3 \right) \delta_p = 2^{y_p} \delta_p
\end{equation}

Where we have defined the exponent $y_p = \log_2(6 - 2 \sqrt{5})$. We now note that we can recover the length scale of our original system via a rescaling by a linear factor of $2$. With this combination of coarse-graining and rescaling, we can generate a scaling relationship for the correlation length $\xi$:

\begin{equation}
\xi(\delta p) = 2 \xi(2^{y_p} \delta_p) \propto p^{\frac{-1}{y_p}}
\end{equation}

Leaving us with the critical exponent

\begin{equation}
\nu = \frac{1}{y_p} = \frac{1}{\log_2(6 - 2 \sqrt{5})} \approx 1.63528
\end{equation}

The final critical exponent we need an independent calculation of is the fractal dimensionality $d_f$. To calculate this, we have to keep track of the mass of a cluster as the coarse-graining procedure is applied. Naively, we would simply include the expected mass of a bond $m_b$ in our RG equations. This misses an important effect, however, because as we coarse-grain we remove dangling bonds which don't connect across an entire diamond. These dangling bonds still contribute to the mass of the original cluster, and must be included in our renormalization. We can include them by introducing a second variable, the ``dangling mass'' $m_d$ attached to the end of a broken bond, which would start at zero at the smallest length scale.

\begin{table}
\begin{tabularx}{\linewidth}{Y|Y|Y|Y } 
Diagram & Probability & Multiplicity & Mass \\ \hline\hline
\makecell{\includegraphics[width=0.5\linewidth]{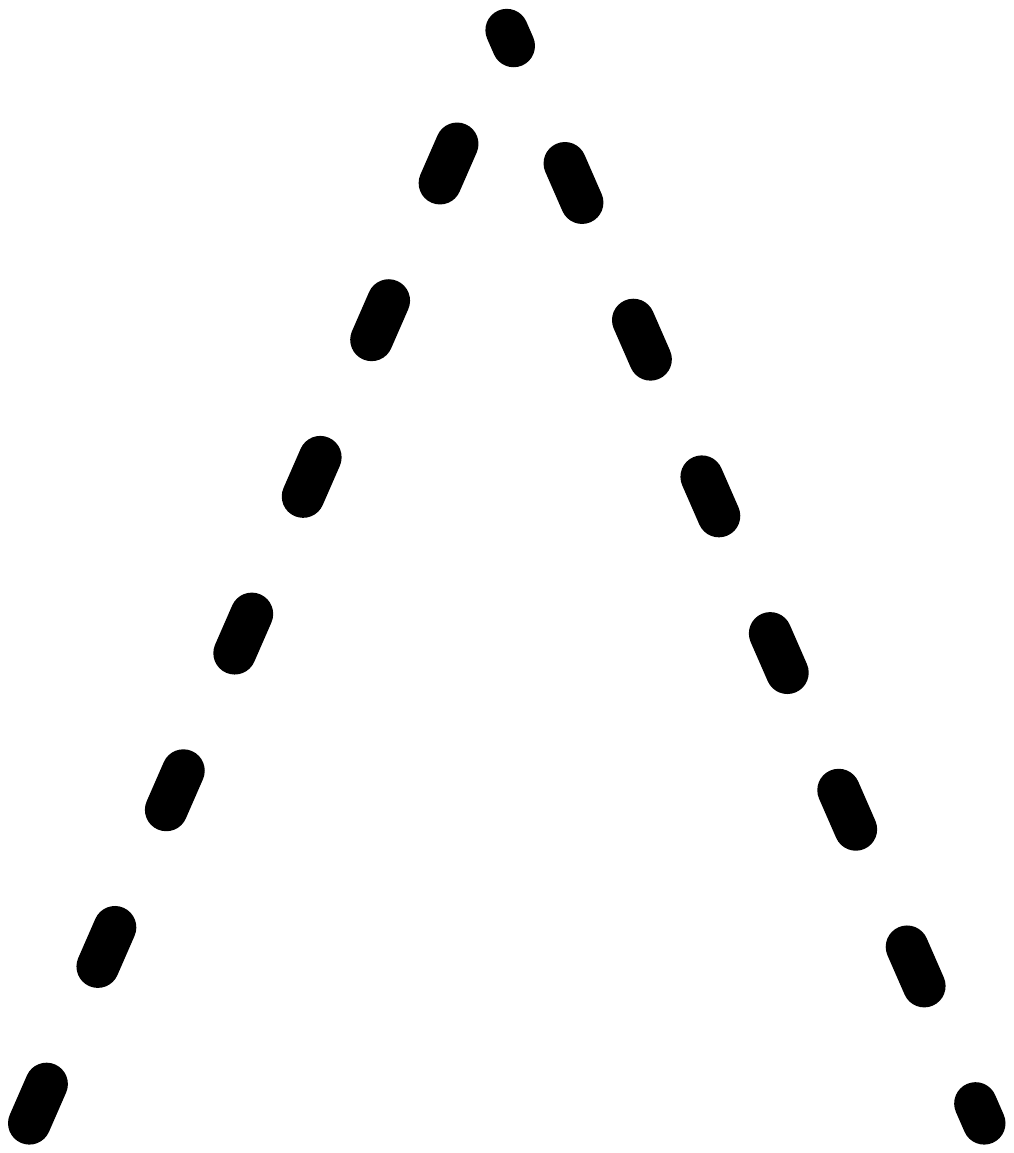}} & $(1-p)^2$ & $1$ & $2 m_d$ \\  \hline
\makecell{\includegraphics[width=0.5\linewidth]{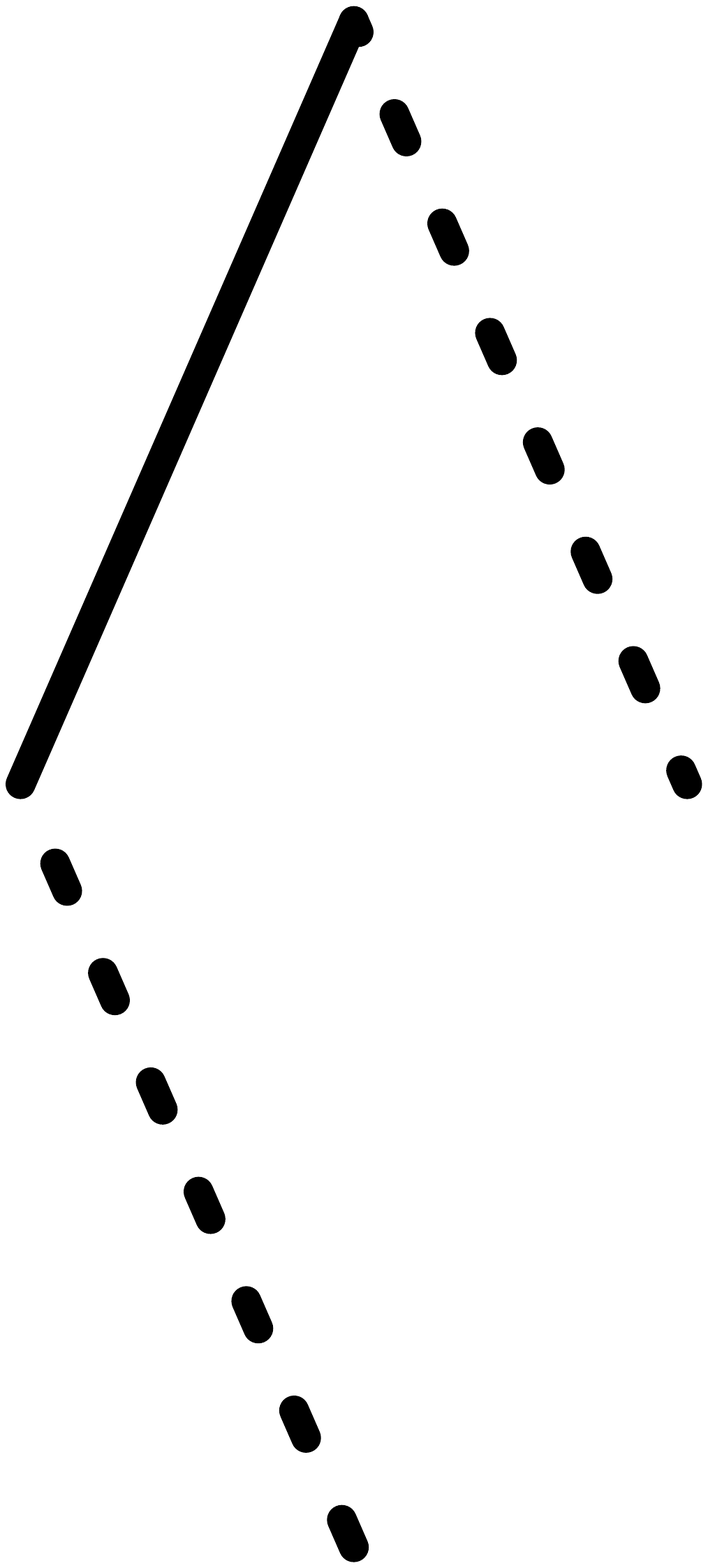}} & $p (1-p)^2 $ & $2$ & $m_b + 2 m_d$ \\ \hline
\makecell{\includegraphics[width=0.5\linewidth]{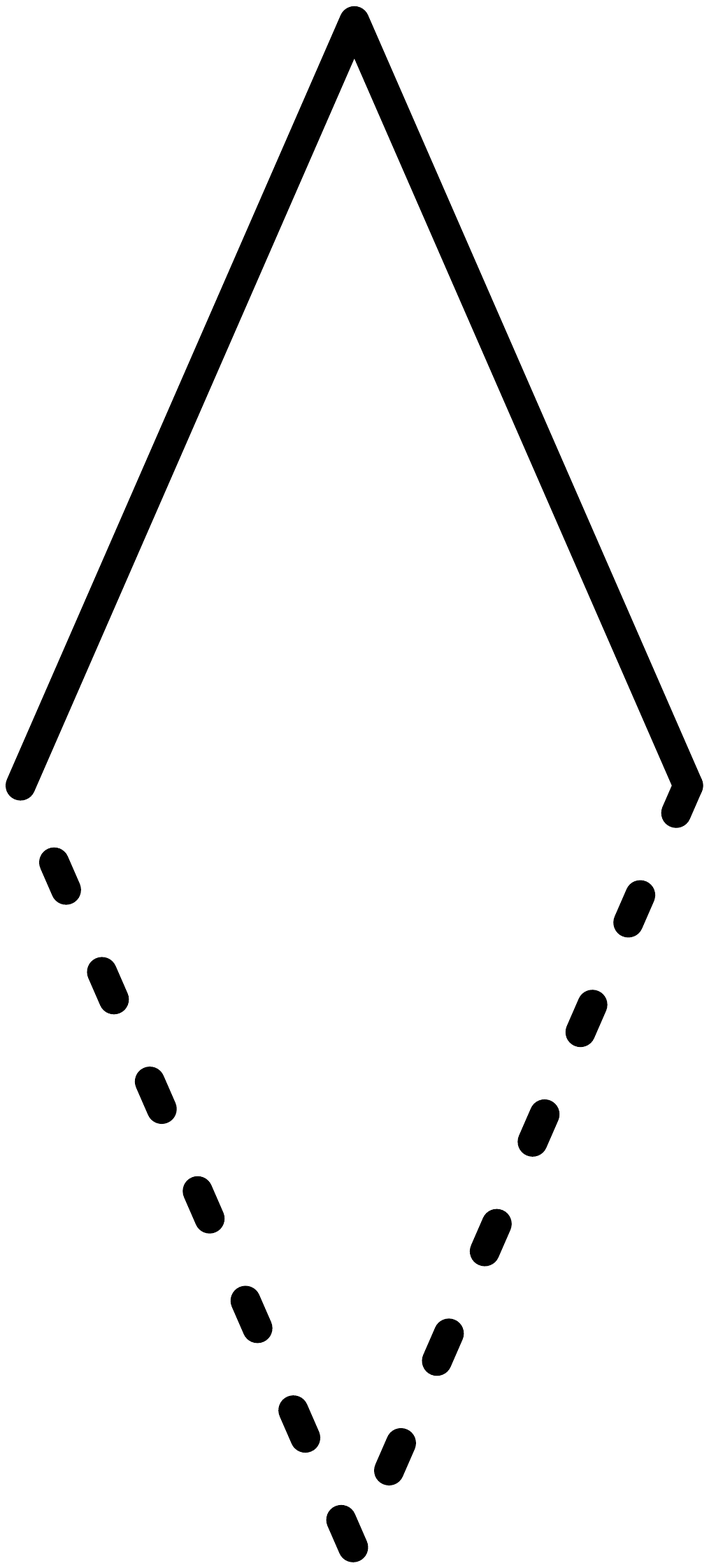}} & $p^2(1-p)^2$ & $1$ & $2m_b + 2m_d$
\end{tabularx}
\caption{The contributions to the renormalized dangling mass $m_d$, calculated for the top vertex (the bottom vertex must obey the same equations by symmetry). Empty bonds represent bonds whose occupancy is irrelevant The probabilities must be normalized by the overall probability $1-p'$ that the coarse-grained bond is broken.}\label{tab:md}
\end{table}

\begin{table}
\begin{tabularx}{\linewidth}{Y|Y|Y|Y } 
Diagram & Probability & Multiplicity & Mass \\ \hline\hline
\makecell{\includegraphics[width=0.5\linewidth]{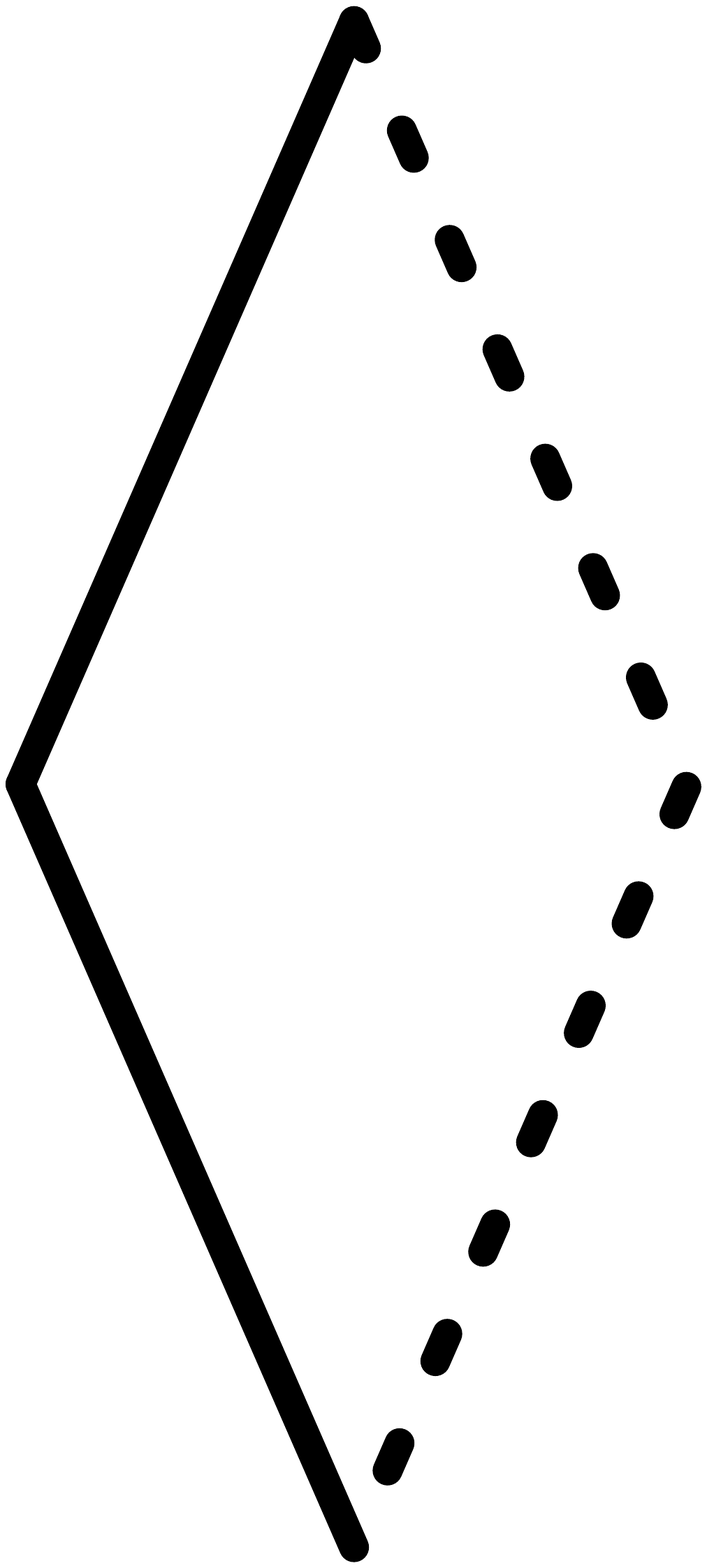}} & $p^2(1-p)^2$ & $2$ & $2 m_b + 2 m_d$ \\ \hline
\makecell{\includegraphics[width=0.5\linewidth]{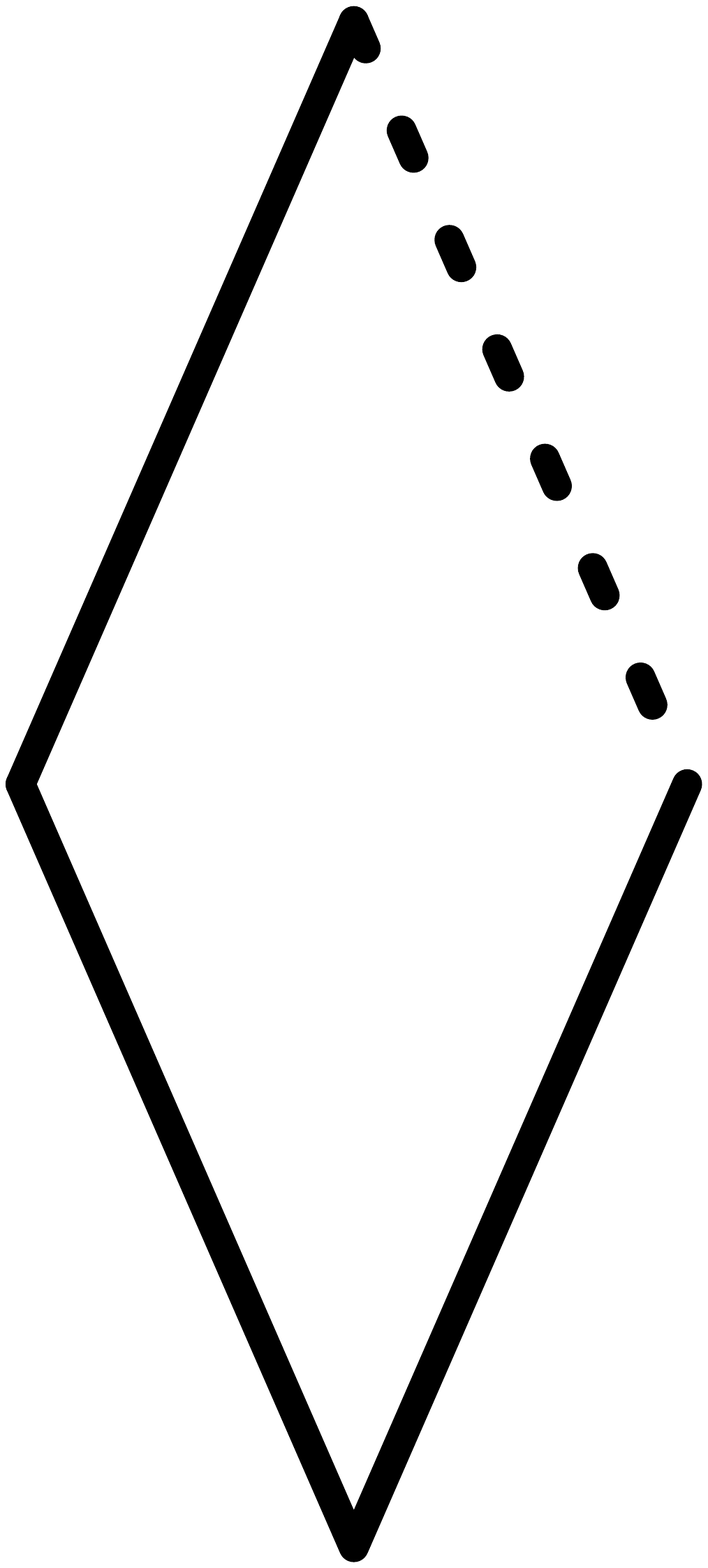}} & $p^3(1-p)$ & $4$ & $3 m_b + 2 m_d$ \\  \hline
\makecell{\includegraphics[width=0.5\linewidth]{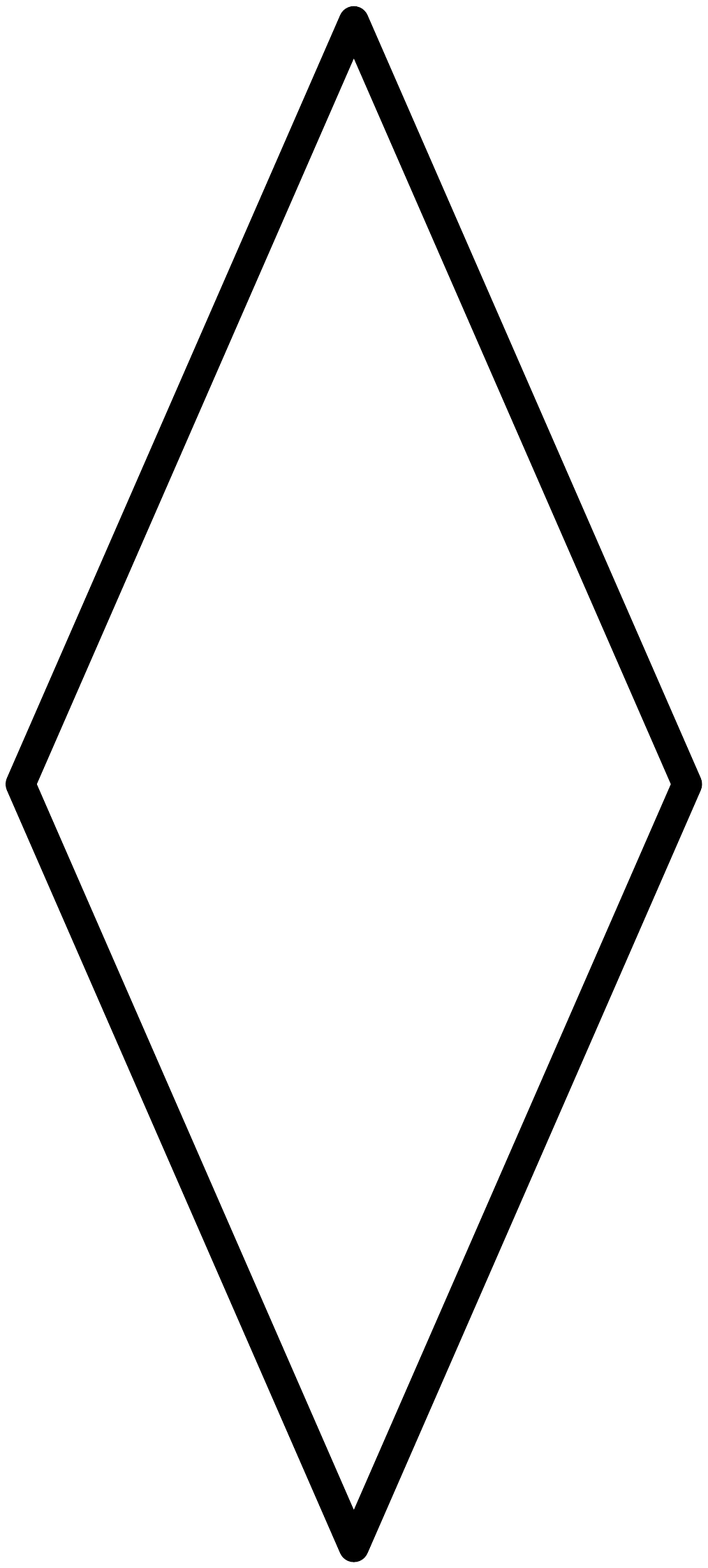}} & $p^4$ & $1$ & $4 m_b$
\end{tabularx}
\caption{The contributions to the renormalized bond mass $m_b$. The probabilities must be normalized by the overal probability $p'$ that the coarse-grained bond exists.}\label{tab:mb}
\end{table}

Formally, we define $m_b$ as the expectation value of the mass of a coarse-grained bond, given that the coarse-grained bond is occupied. $m_d$ is defined as the expectation value of the mass of all the dangling bonds connected to the node at each side of a bond, given that the bond is unoccupied. We can calculate how $m_b$ and $m_d$ update under coarse-graining using tables \ref{tab:mb} and \ref{tab:md} respectively, leading to:

\begin{align}
m_b' &= \frac{1}{p'} \Big[4p^2(m_b + m_d) + 4p^3 m_b - 4p^4(m_b + m_d)\Big] \\
m_d' &= \frac{1}{1-p'} \Big[2 m_d + 2 p m_b - p^2 (2 m_b + 4m_d) \nonumber \\
&\qquad\qquad\ - 2p^3 m_d + 2p^4 (m_b + m_d) \Big]
\end{align}

Where these equations are valid for any $p$. At criticality, $p_c=p'=p$, and these equations reduce to:

\begin{equation}
\begin{pmatrix}m_b' \\[6pt] m_d' \end{pmatrix} = \begin{pmatrix} 4\left(3-\sqrt{5}\right) &\ 2\left(3-\sqrt{5}\right) \\[6pt] \left(3-\sqrt{5}\right) &\ 2 \end{pmatrix} \begin{pmatrix}m_b \\[6pt] m_d \end{pmatrix}
\end{equation}

As we coarse grain the system repeatedly, we expect the distribution of $m_b$ and $m_d$ to converge to the top eigenvector of this matrix, and we associate the scaling of the cluster mass with the top eigenvalue

\begin{align}
\lambda_1 &= 7 - 2\sqrt{5} + \sqrt{73 - 32\sqrt{5}} = 2^{d_f} \nonumber \\
d_f &= \log_2\left(7 - 2\sqrt{5} + \sqrt{73 - 32\sqrt{5}}\right) \approx 1.89929
\end{align}

Knowledge of $\nu$ and $d_f$, together with the underlying lattice dimensionality $d=2$, allow us to calculate all the other geometric critical exponents. The numerical values of the exponents derived from $\nu$ and $d_f$ are summarized in Table \ref{tab:critexp} and compared to the corresponding exponents for standard 2D lattices.

\begin{table}[!ht]
\begin{tabularx}{\linewidth}{Y|Y|Y|Y|Y|Y|Y } 
& $\alpha$ & $\beta$ & $\gamma$ & $\nu$ & $\sigma$ & $\tau$  \\ \hline \hline
2D & -0.667 & 0.139 & 2.389 & 1.333 & 0.396 & 2.055 \\ \hline
Dia & -1.271 & 0.165 & 2.941 & 1.635 & 0.322 & 2.053 
\end{tabularx}
\caption{Comparison of critical exponents in 2D (as compiled in \cite{percflow}) vs on the hierarchical diamond lattice}\label{tab:critexp}
\end{table}

\section{Simulations}

To confirm the plausibility of the RG-calculated exponents, we performed simulations of the system on lattices with depths up to $N=10$ (containing $4^{10}$ bonds). Analysis of the experimental data is limited to checking the consistency of simulation with the presumably exact critical exponents arising from the RG treatment, and we do not systematically extract a set of simulated exponents or confidence intervals.

\begin{figure}
\includegraphics[width=\linewidth]{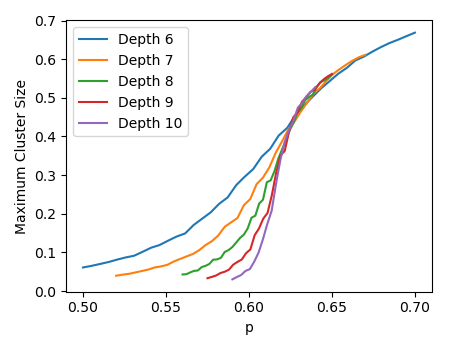}
\caption{The dependence of the order parameter $P$ on the bond occupation probability near the percolation threshold.} \label{fig:pcsims}
\end{figure}

Figure \ref{fig:pcsims} shows the relationship between the order parameter $P$ and $p$ for lattices of various sizes. The theory of finite-size scaling \cite{cardy} implies that all of these curves should be related to a universal curve by the relationship

\begin{equation}
P(p,L) = L^\frac{\beta}{\nu} \tilde{P}((p-p_c)L^\frac{1}{\nu})
\end{equation}

\begin{figure}
\includegraphics[width=\linewidth]{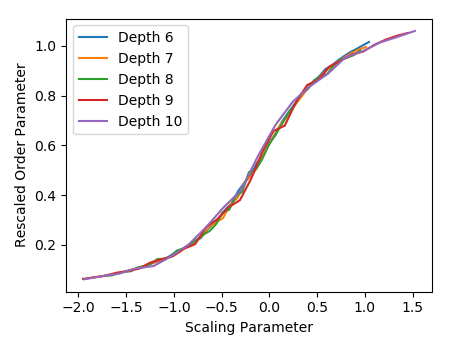}
\caption{Showing the collapse of $P(p)$ near $p_c$ onto a universal curve, using the scaling variables derived from $\nu$ and $\beta$.} \label{fig:fsscaling}
\end{figure}

Where $L = 2^N$ is the characteristic length scale of the overall lattice. Figure \ref{fig:fsscaling} shows that these curves do in fact collapse onto a universal curve in the vicinity of the critical point, using the calculated values for $\beta$, $\nu$ and $p_c$ in the definition of the scaling variables.

\begin{figure}
\includegraphics[width=\linewidth]{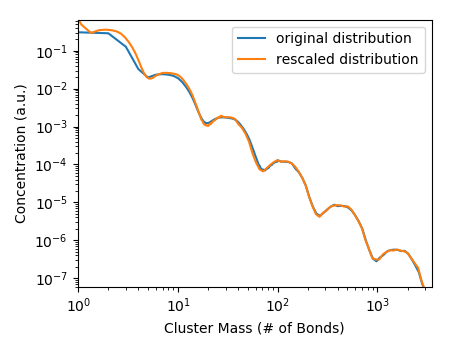}
\caption{The simulated cluster size distribution overlaid with the same distribution after the rescaling operation in \eqref{eq:clustscaling} is applied. The deviation for small clusters is due to the subdominant eigenvalue of the RG equations.} \label{fig:clusterstats}
\end{figure}

Finally, we can probe the cluster statistics $n(s,p_c)$ at the percolation threshold, shown in Figure \ref{fig:clusterstats}. Because our RG equations are defined through an explicit decimation procedure with finite $b=2$, the cluster statistics have an overall decay with an exponent of $-\tau$ modulated by a periodic structure whose periodicity we expect to be $2^{d_f}$. This is related to the fact that the hierarchical lattice has repeating structure on length scales separated by factors of 2. We can apply the scaling transformation

\begin{equation}
n(s,p_c) = 2^d 2^{d_f} n(2^{d_f} s,p_c) \label{eq:clustscaling}
\end{equation}

to confirm the consistency of simulation with the RG result for $d_f$. 

\section{Transport Exponents}

While the geometric exponents are relatively simple to calculate, the transport exponents present a more interesting challenge. The prototypical system which most other transport problems can be reduced to is resistors randomly placed on a lattice. This basic framework can be used to model hydraulic flow through networks, flows of liquid through porous rock or soil, and of course it can be used to study electronic transport in cases such as highly impure semiconductors or actual networks of resistors.

The essential problem involved in performing a RG transformation on the conductances is that even if one begins with only a single type of resistor - say, with a unit conductance, as you coarse grain you wind up with a probability distribution over conductances. When we calculated the fractal dimension of the percolating cluster, the same issue actually exists, but we were justified in ignoring it and treating only the expected value of the masses because the transformations only involved sums of the random variables.

In this case, the mapping from a set of four conductances to a single coarse grained conductance is not just additive. Therefore it is essential to consider distributions over conductances, rather than just the mean conductance, from the outset. This leads to an interesting discussion of how one can apply renormalization group ideas to parameters which are functions, not just numbers.

We define the problem as follows, which parallels the definition in \cite{angulo93} and \cite{stinch76}. With probability $p$, a resistor is placed on each bond. The resistor has a conductance drawn from a distribution $P(\sigma)$ such that $\int d\sigma P(\sigma) = 1$. Our goal now will be to write down a coarse-graining transformation on this probability distribution in addition to a transformation of the probability $p$ (which will necessarily transform the same way as in the geometric percolation problem). Formally, we can write this mapping as:

\begin{align}
p'P'(\sigma) &= 2(p^2-p^4)\int \prod_{i=1}^2 d\sigma_i P(\sigma_i) \delta\left(\sigma - \frac{1}{\frac{1}{\sigma_1} + \frac{1}{\sigma_2}}\right) \nonumber \\
 + p^4 \int &\prod_{i=1}^4 d\sigma_i P(\sigma_i)  \delta\left(\sigma - \frac{1}{\frac{1}{\sigma_1} + \frac{1}{\sigma_2}} - \frac{1}{\frac{1}{\sigma_3} + \frac{1}{\sigma_4}} \right) 
\end{align}

Where we have just applied the general rules for defining probability distributions over functions of independently distributed variables. This can be transformed into the more useful form below, which allows for numerical calculation via a series of two convolutions.

\begin{align}
I(\sigma) &= \int d\sigma' P(\sigma') \left(\frac{1}{\frac{1}{\sigma} - \frac{1}{\sigma'}} \right)^2  P\left(\frac{1}{\frac{1}{\sigma} - \frac{1}{\sigma'}} \right) \\
p'P'(\sigma) &= 2(p^2-p^4)I(\sigma) \nonumber \\ &\quad + (1-2(p^2-p^4)) \int d\sigma' I(\sigma) I(\sigma - \sigma')
\end{align}

At the critical point, $p'=p=p_c$, but this equation is valid in any region of parameter space as long as $p$ is updated at the same time. While it is not clear how to solve analytically for the set of all stationary distributions under this transformation, we can expect to find the conductivity exponent to high precision by iterating this transformation numerically. Ultimately, this procedure is formally equivalent to that defined in \cite{young75} as a first approximation to the RG equations for conductivity on a square lattice. However, in our case, we regard the result as exact (to within computational precision) on the hierarchical diamond lattice.

\begin{figure}
\includegraphics[width=\linewidth]{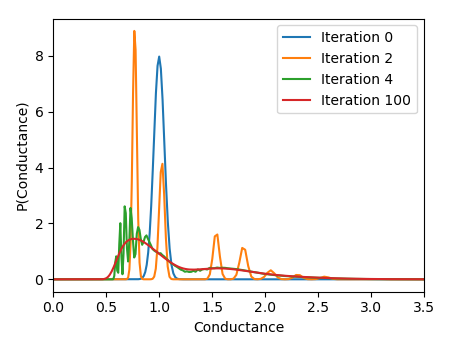}
\caption{Several selected initial steps of the RG equations, starting from a small Gaussian distribution of conductances.} \label{fig:condrescale}
\end{figure}

Figure \ref{fig:condrescale} demonstrates the evolution of an initially tight distribution of conductances as this procedure is iterated. Ultimately, the distribution converges to a smooth bimodal form. In the limiting form, the distribution obeys the scaling equation

\begin{equation}
P'(\sigma) = P(\sigma/\lambda) * \lambda
\end{equation}

With $\lambda \approx 1.75625$ extracted from a regression on the evolution of the distribution's mean. This result implies that the conductance of the distribution scales like $K = L^{\log_2(\lambda)}$ at criticality. When the lattice is near criticality, the medium is homogeneous on length scales larger than $\xi$, implying that the conductivity scales as $\sigma(p) = (p-p_c)^t$ with $t = \nu \log_2(\lambda) \approx 1.32866$.

\section{Conclusion}

In conclusion, we have calculated exact values for the various geometric exponents on the $d_e=2$ hierarchical diamond lattice. The calculated values match results from explicit simulation of the percolation problem to within the resolution of the computational experiments. In addition, we have used a renormalization group operation defined on probability distributions to calculate a numerical value for the conductivity exponent $t$.

\bibliography{references}

\end{document}